\documentclass[fleqn,usenatbib]{mnras}

\usepackage{newtxtext,newtxmath}

\usepackage[T1]{fontenc}

\DeclareRobustCommand{\VAN}[3]{#2}
\let\VANthebibliography\thebibliography
\def\thebibliography{\DeclareRobustCommand{\VAN}[3]{##3}\VANthebibliography}

\usepackage{graphicx}	
\usepackage{amsmath}	
\usepackage{subcaption}
\usepackage[export]{adjustbox}
\usepackage{wrapfig}
\usepackage{longtable}
\usepackage{pdflscape}
\usepackage{geometry}
\usepackage{supertabular,booktabs}
\usepackage{floatrow}
\usepackage{float}
\usepackage{xcolor}
\usepackage{tablefootnote}
\usepackage{threeparttable}

\newcommand{\src}{Aql~X-1}
\newcommand{\nicer}{\textit{NICER}}
\newcommand{\changes}[1]{{\color{black} #1}}
\def\lsim{\mathrel{\rlap{\lower4pt\hbox{\hskip1pt$\sim$}}
    \raise1pt\hbox{$<$}}}         
\def\gsim{\mathrel{\rlap{\lower4pt\hbox{\hskip1pt$\sim$}}
    \raise1pt\hbox{$>$}}}         

\title[Spectral evolution of Aql X-1]{Probing outbursts of the transient neutron star low mass X-ray binary Aql~X-1 with \textit{NICER}: a study of spectral evolution}

\author[Putha et al.]{
Karthik Gananath Putha$^{1}$\thanks{E-mail: kputha@email.sc.edu},
Yash Bhargava$^{2}$ and
Sudip Bhattacharyya$^{2}$
\\
$^{1}$Department of Physics and Astronomy, University of South Carolina, 712 Main St., Columbia 29205, USA \\
$^{2}$Department of Astronomy and Astrophysics, Tata Institute of Fundamental Research, 1 Homi Bhabha Road, Colaba, Mumbai 400005, India\\
}

\date{Accepted XXX. Received YYY; in original form ZZZ}

\pubyear{2023}

\begin{document}
\label{firstpage}
\pagerange{\pageref{firstpage}--\pageref{lastpage}}
\maketitle

\begin{abstract}
X-ray observations of neutron star (NS) low mass X-ray binaries (LMXBs) are useful to probe physical processes close to the NS and to constrain source parameters. Aql X-1 is a  transient NS LMXB which frequently undergoes outbursts provides an excellent opportunity to study source properties and accretion mechanism in strong gravity regime over a wide range of accretion rates.
In this work, we systematically investigate the spectral evolution of Aql X-1 using \textit{NICER} observations during the source outbursts in 2019 and 2020. The \textit{NICER} observations cover the complete transition of the source from its canonical hard state to soft state and back. 
\changes{The spectra extracted from most  observations can be explained by a partially Comptonised accretion disc. We find that the system can be described by an accretion disk with an inner temperature of $\approx0.7$~keV and a Comptonising medium of thermal electrons at $\approx2$~keV, while the photon index is strongly degenerate with the covering fraction of the medium.  We also find evidence of Fe K$\alpha$ fluorescence emission in the spectra indicating reprocessing of the Comptonised photons. We observe an absorption column density higher than the Galactic column density for most of the observations indicating a significant local absorption. But for some of the observations in 2020 outburst, the local absorption is negligible.  
}

\end{abstract}

\begin{keywords}
accretion, accretion discs---methods: data analysis---stars: neutron ---X-rays: binaries---X-rays: individual (Aql X-1).
\end{keywords}



\section{Introduction}


Low mass X-ray binaries (LMXB) consists of a neutron star (NS) or a black hole (BH) and a companion star with a mass $\lesssim$ 1~M$_{\odot}$. In these binaries, the compact object accretes the matter from the companion star through the Roche-Lobe overflow. The NS LMXBs are broadly categorised into `atoll' sources or `Z' sources based on the shape of the track generated on their hardness intensity diagram (HID) or the colour-colour diagram (CCD) \citep[see][for a review]{vdkReview2004astro.ph.10551V}. 
Most of the NS LMXBs are transient systems which stay in the quiescent phase for long duration and undergo outbursts during which they are typically detected. 
These systems produce outbursts, due to the increased mass accretion rate, which are caused by the instabilities in the disc \citep{SS1973A&A....24..337S, Done2007A&ARv..15....1D}. During these outbursts, the luminosity of the system increases and it emits photons from the radio to the X-ray band \citep{Done2007A&ARv..15....1D}.

The X-ray emission from NS LMXBs has been modelled with a combination of spectral components. The primary component of the modelling is the accretion disc \citep{SS1973A&A....24..337S}, a blackbody component (from the accretion mound on the surface of the NS or from the boundary layer) and a non-thermal emission (either from the accretion column or the Comptonised emission). The Comptonised emission results from Compton up-scattering of the soft thermal photons from the accretion disc by a hot plasma of electrons (often called corona) close to the NS. During the evolution of the source across various accretion states, there is a strong interplay between these components which decides the track followed by the source on the HID/CCD.


Aquila X-1 \citep[or \src, discovered by][]{kunte1973NPhS..245...37K} is one of the most famous NS LMXBs which regularly goes into an outburst every year \citep[although the outburst is not exactly periodic,][]{simon2002A&A...381..151S, ootes2018MNRAS.477.2900O}. Due to the extensive observations of the source in various wavelengths, key parameters of the source are well known. \src\ has a K-type donor and is placed at a distance of $6\pm2$~kpc \citep{mata2017MNRAS.464L..41M}. The orbital period of the binary system is 18.95 d \citep{chevalier1991A&A...251L..11C, welsh2000AJ....120..943W}.   The presence of the thermo-nuclear bursts in the X-ray light curves suggest the compact object is a NS \citep[e.g. ][]{galloway2008ApJS..179..360G, guver2022MNRAS.510.1577G} and the detection of the millisecond pulsations in one such bursts \citep[1.8~ms,][]{casella2008ApJ...674L..41C} confirms the nature of the accretor. 
\src\ is one of the few NS LMXBs which shows a hysteresis in the HID \citep[][]{Maitra2004ApJ...608..444M, guver2022MNRAS.510.1577G} similar to BH LMXBs \citep[e.g.][and references within]{bhargava2019MNRAS.488..720B, rawat2023MNRAS.520..113R}. 
Using \textit{RXTE}-PCA observations of \src,  \citet{Maitra2004ApJ...608..444M} have studied the evolution of the source across the outburst in 2000 as it traverses various spectral states i.e. low hard state (LHS) and high soft state (HSS) and other intermediate states. The authors characterise the spectrum of the source as a combination of an accretion disc and a powerlaw component. Similar characterisation of the source evolution across different types of the outbursts (i.e. long high, medium low and short low) was conducted using monitoring observations of \src\ with \textit{RXTE}/PCA by \citet{gungor2014MNRAS.439.2717G}. The long high outbursts are more frequent and during the majority of the outburst, the source stays in the HSS while tracing a hysteresis across the other states during the rise and the decay phase of the outburst \citep{Maitra2004ApJ...608..444M}. \citet{lopez2020MNRAS.493..940L} have undertaken preliminary investigation of the soft X-ray evolution of the source using \textit{Swift}-XRT spectra but mainly focus on the correlation of the UV-optical properties and their correlation with the X-ray spectral evolution. Recently, \citet{niwano2023MNRAS.525.4358N} studied outbursts of the source from 2016--2019 using the \textit{MAXI} instrument onboard ISS where they model the 2--20~keV spectrum as the emission from an absorbed accretion disc. 

The soft X-ray evolution of NS LMXB holds the key to evolution of the thermal components \citep[e.g.][]{bhargava2023arXiv230713979B}. In case of \src, the soft X-ray spectrum is somewhat understudied. 
In this article, we investigate the properties of \src\ in soft X-rays using extensive \nicer\ monitoring during its outbursts in 2019 and 2020. We report the details of the observation and data reduction methods in section~\ref{sec:obs}, spectral analysis methodology in section~\ref{sec:spec} and describe the results and our interpretation in section~\ref{sec:res}.

\section{Observations and Data reduction}\label{sec:obs}

\nicer\ \citep{Gendreau2016SPIE.9905E..1HG} has regularly observed \src\ since its launch in 2017. Since then \src\ has undergone several outbursts. For this analysis, we consider the outbursts that happened from 2019 August 2 - 2019 September 21 (outburst A) and from 2020 August 15 - 2020 October 23 (outburst B). These observations are considered for the analysis as they are representative of the complete outburst (from the rise to the decay).  The details of the observations are reported in tables~\ref{tab: Outburst A log table}  and \ref{tab: Outburst B log table}.  In the draft, we refer to the observations with A/BXXX, where A/B denote which outburst they are part of and the XXX indicates the last three digits of the observation id (ObsID) as mentioned in tables~\ref{tab: Outburst A log table}  and \ref{tab: Outburst B log table}. 

\changes{The data is processed through a standard filtering pipeline, using \texttt{nicerl2} i.e, excluding the detectors 14, 34 and excluding the time intervals corresponding to South Atlantic Anomaly (SAA) passages,  the Earth elevation angle $>$ $15^{\circ}$,  the elevation angle with respect to the bright Earth limb $\lesssim 30^{\circ}$, and considering the intervals where the undershoot and overshoot count rates per module are 0-500 and 0-30 respectively}. We used \textsc{HEASOFT} V6.31, \textsc{NICERDAS} V10a and CALDB version xti20221001.

\begin{figure*}
    \centering
    \includegraphics[width=0.49\textwidth]{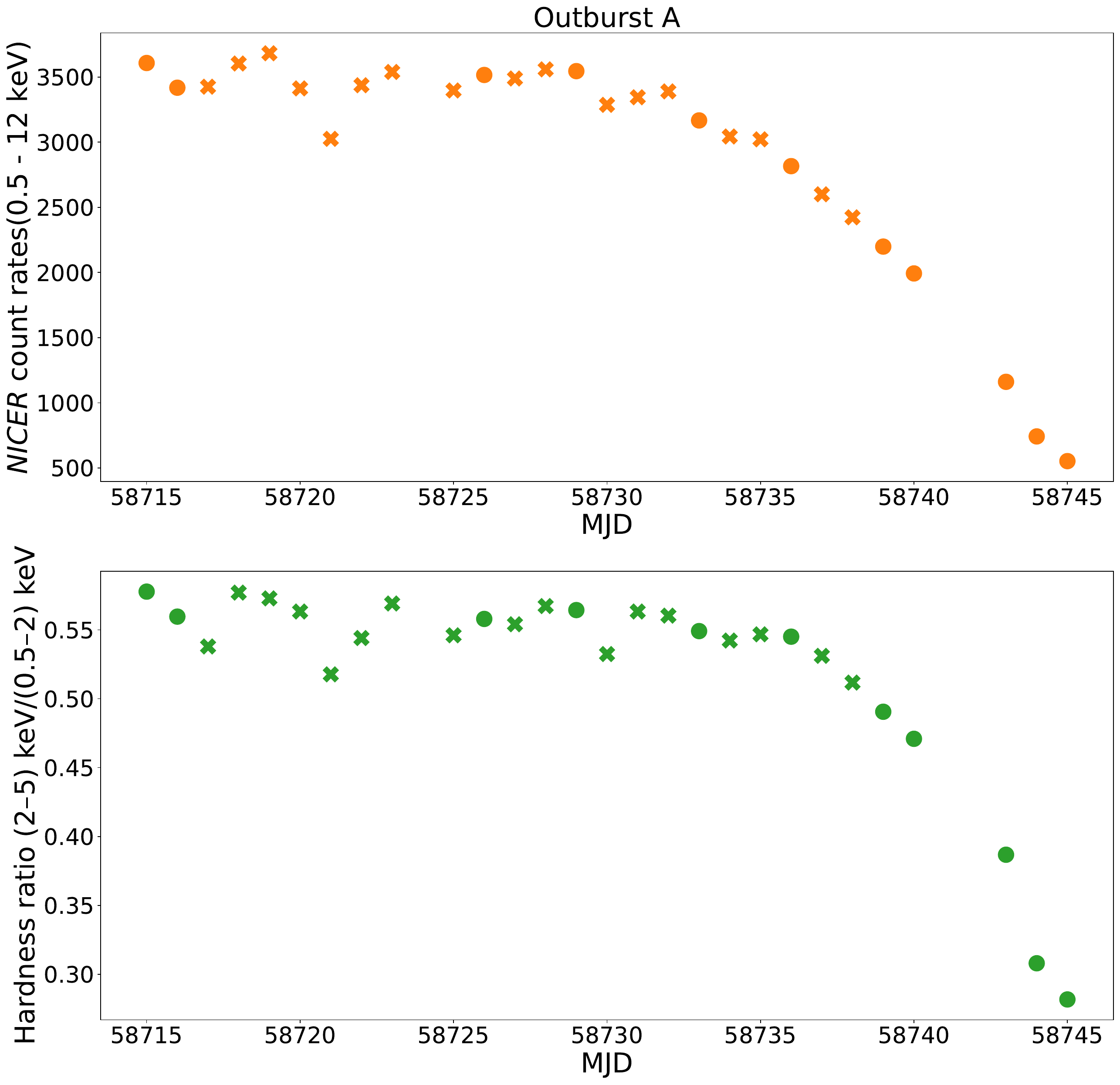} \includegraphics[width=0.49\textwidth]{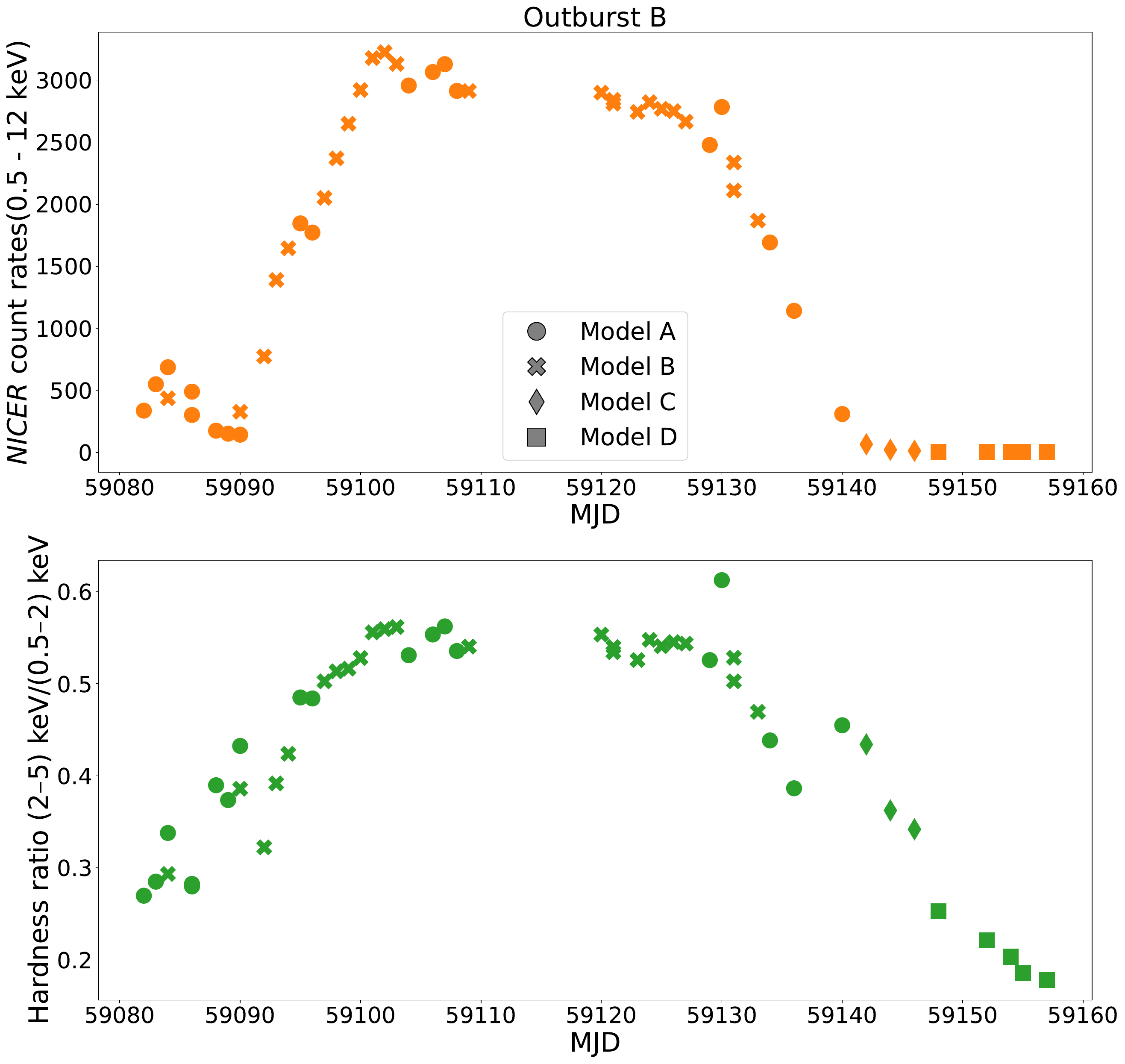}
    
    \caption{\changes{The plots have been updated} Outburst of \src\ in 2019 (left panels)  and 2020 (right panels). The top panels (both left and right) show the evolution of the count rate in 0.5-12~keV while the bottom panels show the evolution of the hardness ratio in 2-5/0.5-2~keV energy bands. Each point corresponds to a single observation. \changes{The shape of the points have been assigned according to the spectral decomposition,i.e, Model A = \texttt{tbabs*(gauss+thcomp$\otimes$diskbb)} by a circle, Model B = \texttt{tbabs*(gauss+gauss+thcomp$\otimes$diskbb)} by a cross, Model C = \texttt{tbabs*(thcomp$\otimes$diskbb)} by a diamond, and Model D = \texttt{tbabs*(diskbb)} by a square (see the text for a detailed description of the models). } 
    }
    \label{fig:lc_hr}
\end{figure*}

\begin{figure*}
    \centering
    \includegraphics[width=0.8\columnwidth]{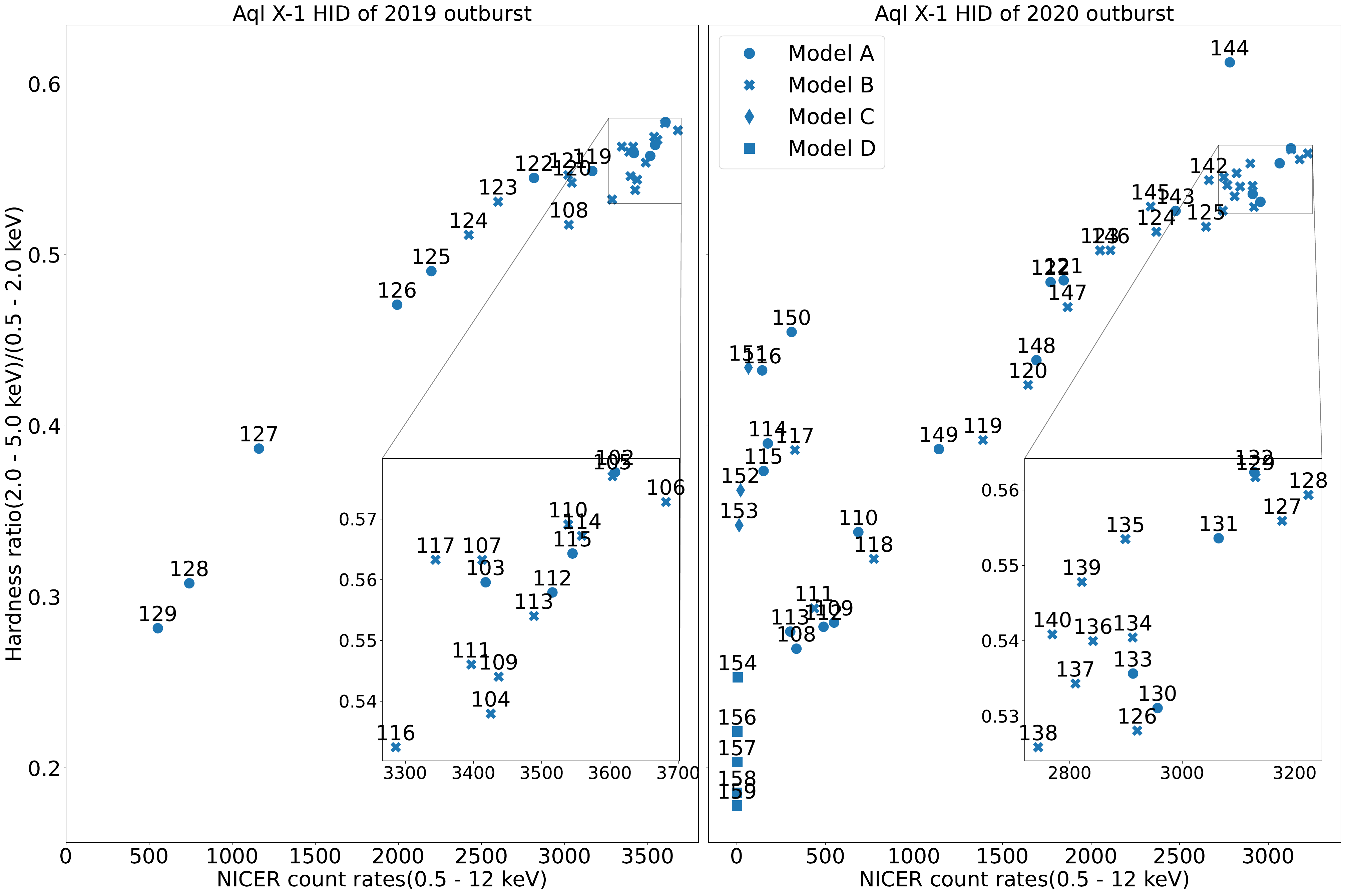}
    \caption{\changes{The plot has been updated} Hardness intensity diagram of 2019 (left panel) and 2020 (right panel) outbursts of \src. Hardness ratio is defined as the ratio between the  2--5 and 0.5--2~keV light curve is plotted against the intensity in 0.5--12~keV energy range. The inset plot shows a better view of the clustered points in the HSS. The marker scheme is same as Figure~\ref{fig:lc_hr}.  
    }
    \label{fig:hid}                                               
\end{figure*}

Using \texttt{nicerl3-lc}, light curves are extracted in the energy ranges 0.5--12.0~keV, 0.5--2.0~keV, and 2.0--5.0~keV. For further analysis, we ignore the observations with the net exposure $\lesssim$ 200~s. We depict the light curves in 0.5--12~keV in the top panels of figure~\ref{fig:lc_hr} and the hardness ratio of computed using 2--5~keV and 0.5--2~keV light curves in the bottom panels of figure~\ref{fig:lc_hr}. We also construct a HID of the source for both the outbursts in figure~\ref{fig:hid}. Some of the observations analysed here show bursts in their light curve.  \changes{Since we want to focus on the persistent emission of the source, we identify the time stamps of the bursts from the light curve using \texttt{xselect} and create Good Time Intervals (GTIs) excluding the bursts\footnote{\changes{\url{https://swift.gsfc.nasa.gov/analysis/threads/batlightcurvethread.html}, is used to construct GTI}}. Information about these bursts are listed in table \ref{tab: Bursts table}. To exclude the burst, we remove the interval during which the source exceeded the mean count rate before and after the burst, without assuming any decay timescale. These GTIs are then passed to  \texttt{nicerl3-spec} routine while extracting the spectrum. }  
The burst intervals have been excluded from the light curves plotted in the above-mentioned figures and the further spectral analysis. 

\section{Spectral Analysis}\label{sec:spec}

\begin{figure}
    \centering
    \includegraphics[width=\columnwidth]{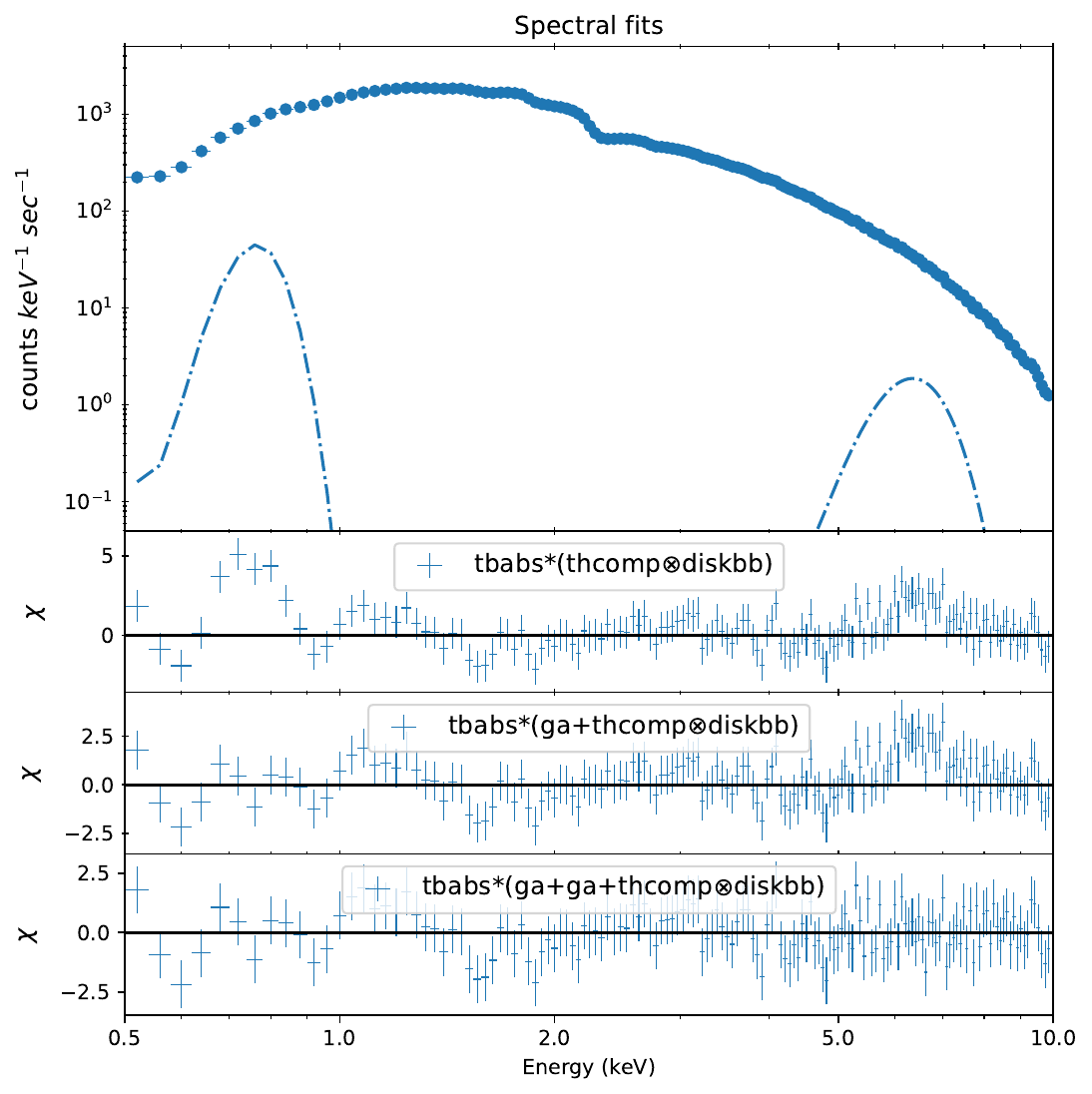}
\caption{\changes{Count spectra and spectral decomposition (top panel), and residuals ($\chi=$ (Data - Model)/Error)  without both Gaussians (second panel), without Gaussian for iron line (third panel) and with both Gaussians (bottom panel) for A116 observation with the model B having a fixed covering fraction \textit{f}=0.75.}}


    \label{fig:spectral fits}
\end{figure}

To investigate the spectral evolution of \src, we characterise the spectrum as seen by \nicer\ in \changes{0.5--10~keV} for all the observations. For modelling the spectrum, we use the $\chi^2$-statistic and report the 1$\sigma$ uncertainty in the parameters throughout the article.  
We use the SCORPEON model \citep{nicer2023HEAD...2010349M} to estimate the background, and fix the background model parameters to the default values. 
In case of the fainter observations (count rate in 0.5--12~keV $\lesssim100$ cts/s, B149-B159), we restrict the spectrum to the energy ranges where the source counts are above the expected background level. 
\changes{ A systematic error of 0.8\% is used during the spectrum modelling in the operational energy range of 0.5--10~keV. We have used the optimal binning by the spectral response  \citep{2016A&A...587A.151K} for the spectral modelling.
We have depicted spectrum of observation A116 in the top panel of the figure~\ref{fig:spectral fits}. }

To test the nature of the spectrum, we model the spectra from the individual observations with, a single additive component (viz. \texttt{diskbb}, \texttt{bbodyrad} or \texttt{nthcomp}), along with an absorption model to correct for the interstellar+intrinsic absorption (modelled with \texttt{tbabs}).  The abundances from \citet{Wilms2000ApJ...542..914W} and the cross-sections from \citet{Vern1996ApJ...465..487V} are used to model the spectrum. 
\changes{The single-component model is found to be insufficient to explain the spectrum with count rate $>$100 counts. For example, ObsID A121 the
model \texttt{tbabs*(diskbb)} result in $\chi^{2}$/degrees of freedom (DoF)
of 8852.37/158, but on inclusion of Comptonisation model \texttt{thcomp} \citep{thcomp2020MNRAS.492.5234Z} result in $\chi^{2}$/DoF of  144.22/156. 
The drastic change in $\chi^{2}$ for a minimal change in the DoF indicates a strong presences of a Comptonized emission.  
Thus we model all the spectra at least as a combination of a thermalised accretion disc and a Comptonised emission i.e, \texttt{tbabs*(thcomp$\otimes$diskbb)}. Across all ObsIDs (except the fainter observations, i.e. B151--B159), we identify residuals around 0.75~keV which can be modelled by a Gaussian. This feature has also been seen in other sources with \nicer\ indicating that it is an instrumental feature \citep[e.g.][]{2023MNRAS.526.3944Z}. For example, in A112, adding a Gaussian component reduces $\chi^{2}$/DoF from 191.97/143 to 153/140, resulting in a f-test statistic of 11.88 and probability of 5.55$\times10^{-7}$. Thus we model the observations by including the Gaussian in the model (\texttt{tbabs*(gauss+thcomp$\otimes$diskbb)}; Model A).} 
\changes{The residuals of some spectra indicate the presence of a weak Fe K$\alpha$ emission line which is modelled by a Gaussian, i.e \texttt{Tbabs*(gauss+gauss+thcomp$\otimes$diskbb)} (Model B). For example, the addition of Gaussian for A116 (which shows a structured residual near 6.5~keV with Model A) results in a f-test value of 22.55 with probability 2.97$\times10^{-9}$, whereas adding a Gaussian to A112 (with no significant residual near 6.5~keV with Model A) results in an f-test value of 2.84 with probability 0.061. Some observations at the end of outburst B, i.e, from B151-B153 do not show evidence for the instrumental Gaussian component and are modelled by \texttt{tbabs*(thcomp$\otimes$diskbb)}; (Model C) and fainter observations B154-B159 are modelled by \texttt{tbabs*(diskbb)} (Model D)
}. \changes{We depict the spectrum and its spectral decomposition of A116 in the \changes{first} panel of figure~\ref{fig:spectral fits} and the residuals to the spectral fits in the bottom panels of figure~\ref{fig:spectral fits}. The second panel shows the residuals with both Gaussians, third panel shows the residual without the Gaussian at 6.5~keV and last panel displays the residual with out Gaussian at 0.75~keV.} 


The convolution model \texttt{thcomp}  up-scatters a fraction of the seed photons ($f$) from the disc and returns a combination of the un-scattered disc emission and Comptonised emission (of a photon index $\Gamma$ and electron temperature $kT_e$). \changes{Since convolution models requires computation of the model beyond the spectral fitting range, we used a custom energy grid from 0.01--100~keV using 1000 logarithmic bins.}

\begin{figure}
    \centering
    \includegraphics[width=\columnwidth]{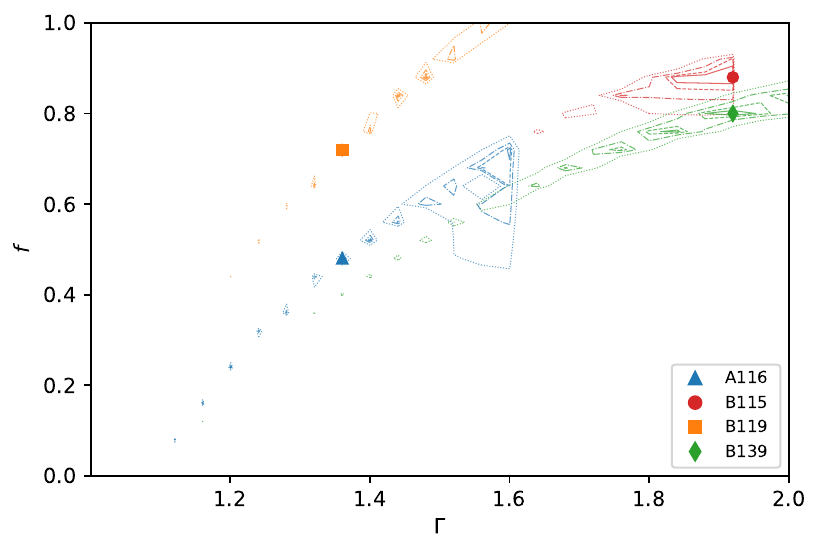}
\caption{\changes{ $\chi^2$ contours of observations from different parts of HID, displaying the degeneracy between the parameters $\Gamma$ and \textit{f} are depicted in different colours. The minimum of $\chi^{2}$ for each observation are indicated by different markers (and the symbol is included in the legend).  We depict  the contours corresponding to 68$\%$ (solid), 90$\%$ (dashed), 99$\%$ (dashed-dot) and 99.99$\% $(dotted) confidence levels.  }
}

    \label{fig:contour}
\end{figure}

\changes{If we keep all parameters free across the different observations, we find that some of the observations indicated a low $\Gamma$ ($\sim$1) while some required high $\Gamma$ ($\sim$1.8), while they lie at a similar HID position. These observations also show low and high $f$ respectively. To probe if there is a degeneracy between these parameters, we generate $\chi^2$ contour maps by sampling a uniform grid of $\Gamma$ from 1.001 to 2 and $f$ from 0 to 1 in the steps of 25. We find that for all observations these two parameters are degenerate but the degeneracy curve depends on the source position on the HID. We depict the contour plot for some of the observations from different parts of the HID in figure \ref{fig:contour}. }

\begin{figure*}
    \centering
    \includegraphics[width=0.8\columnwidth]{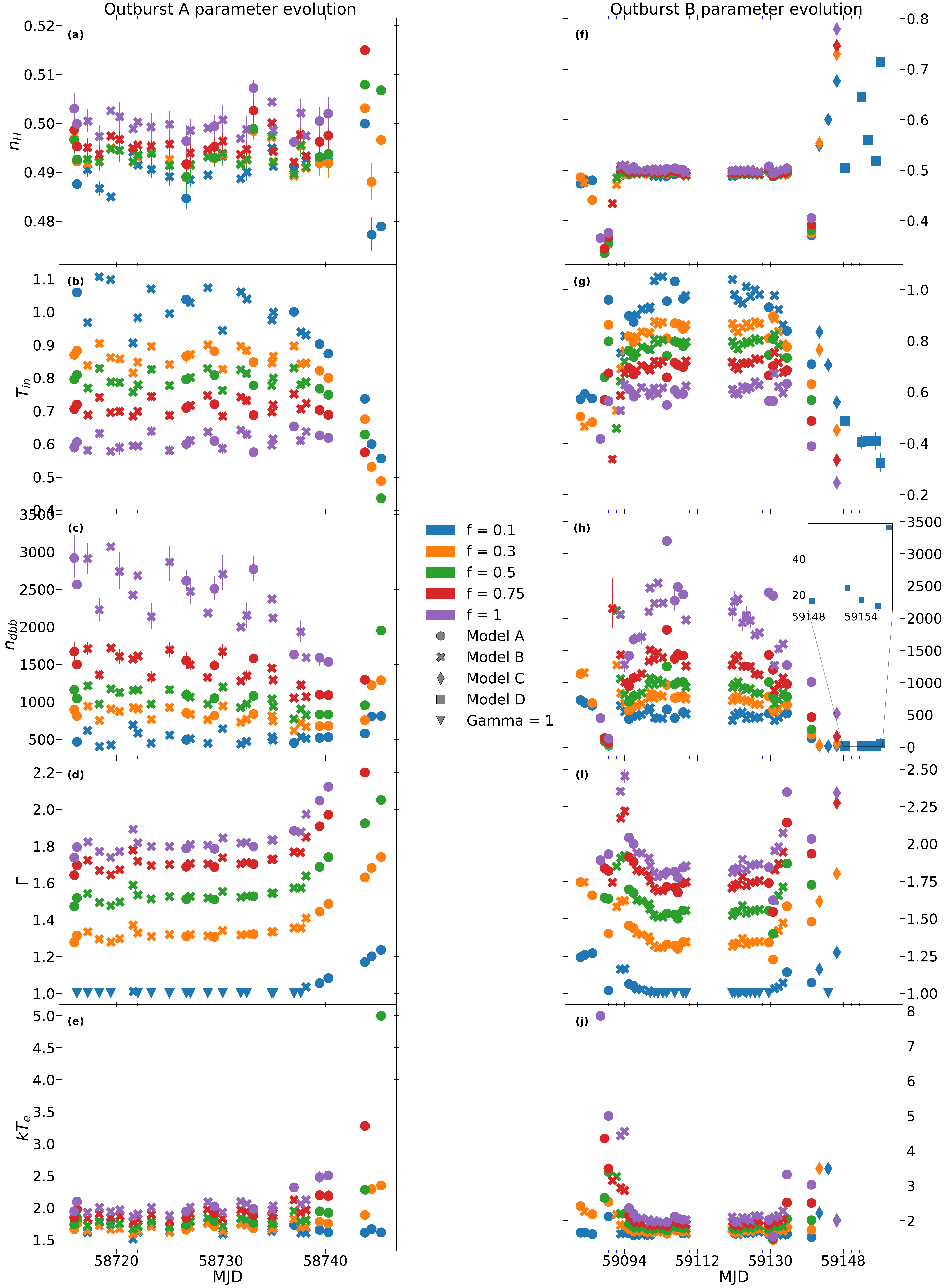}
    \caption{ \changes{Spectral parameter evolution of Outburst A (Right panels) and B (left panels). Only the parameters of the observations with $\chi^{2}$/DoF$<1.5$. Different colours are used to show the evolution of parameters for different $\textit{f}$. The marker shapes are kept identical to figure \ref{fig:lc_hr}. The inverted triangles in panels (d) and (i) indicate the observations which peg of $\Gamma$=1. The inset plot in panel (h) shows the disc norm evolution of the fainter observations in outburst B.The parameter values for the outbursts for $f=0.75$ are noted in tables~\ref{tab:2019 Outburst Parameters} and \ref{tab:2020 Outburst Parameters}}. 
    }
    
    \label{fig: Parameter evolution}
\end{figure*}

\changes{Due to a degeneracy in the parameter space, we cannot probe both parameters simultaneously. To understand the parameter evolution and observe the trends in different parameters as a function of time, we fix the $f$ to a range of values 0.1, 0.3, 0.5, 0.75, and 1.0, and estimated the spectral parameters for all the observations. The evolution of the parameters is shown in figure~\ref{fig: Parameter evolution}. We note that not all observations result in acceptable fits (indicating a $\chi^2$/DoF $>$ 1.5) on fixing $f$ and the parameters of unacceptable fits are not shown in figure~\ref{fig: Parameter evolution}. We report the parameters for $f=0.75$ in tables \ref{tab:2019 Outburst Parameters} and \ref{tab:2020 Outburst Parameters} as most of the observations have a $\chi^{2}$/DoF $<$ 1.5.}




\changes{We compute the unabsorbed flux in the energy range of 0.5--12.0~keV using the convolution model \texttt{cflux}. We use the same energy grid as we did for the \texttt{thcomp}. For computing the flux, we freeze all the parameters and then add the \texttt{cflux} component, fitting the model. Then refit the model by thawing the non-normalisation parameters (e.g. $T_{\rm in}$) while keeping the Gaussian and \textit{f} frozen to get the flux value. The flux remains the same irrespective of \textit{f}. We report the unabsorbed flux in Tables~\ref{tab:2019 Outburst Parameters} and \ref{tab:2020 Outburst Parameters}.  
}


\section{Results and discussion}\label{sec:res}

We investigated the evolution of the continuum properties of the LMXB \src\ in the soft X-rays using the \nicer\ observations of the outbursts of the source in 2019 and 2020 (Outburst A and B respectively). The outbursts covered the complete evolution of the source and revealed an interesting trend in the spectral properties of the source.

\subsection{Light curves and HID}

The \nicer\ observations of \src\ in 2019 and 2020 indicate an evolution of the source which had been extensively seen in the literature \citep{Maitra2004ApJ...608..444M, gungor2014MNRAS.439.2717G, niwano2023MNRAS.525.4358N}. The outbursts of the source investigated here (see figure~\ref{fig:lc_hr}) are the `long high' kind with the profile resembling typical fast rise exponential decay \citep[FRED,][]{Chen1997ApJ...491..312C}. Such outbursts have been seen often in \src, indicating that these outbursts are typical of the source. 

\src\ is known to show a Q-shaped hysteresis in its HID \citep[using the hardness of 9.7--16/6.4-9.7~keV from \textit{RXTE}/PCA, e.g. ][]{Maitra2004ApJ...608..444M} as the source evolves in the outburst. In the case of the hardness defined as 9.7--16/6.4-9.7~keV, the count rate and the hardness typically show an anti-correlation (for a higher count rate, a lower hardness was observed and vice versa).   The track changes its shape when a softer definition of the hardness was used \citep[3.8--6.8/2--3.8~keV using \nicer,][and 2--5/0.5--2~keV figure~\ref{fig:hid}]{guver2022MNRAS.510.1577G}, with a change in the trend between the count rate and hardness. For the softer definition of hardness, the count rate was proportional to hardness for the majority of the observations. The trend between the hardness and intensity depends mainly on the spectral shape of the source. A positive trend indicates that with increasing count rate, the harder band has more photons than the softer band and vice versa. In case of \nicer\ observations, the harder band was 2--5~keV in the current work and 3.8--6.8~keV in \citet{guver2022MNRAS.510.1577G} while the softer band was 0.5--2~keV and 2--3.8~keV respectively. Since in both of these cases, we see a positive trend between hardness and intensity, it indicated that the count rate was increasing by a larger amount in the bands 2--6.8~keV as compared to 0.5--2~keV. And the negative trend for a similar change in the count rate in 9.7--16/6.4--9.7~keV from \textit{RXTE}/PCA observations indicated in 6.4--9.7~keV band the count rates were increasing more as compared to 9.7--16~keV. Thus, in a model-independent way, we ascertain that during an outburst, there was a strong spectral change, with $\approx$ 2--6~keV count rate changing with the total count rate more strongly as compared to the energy bands around it.

The Q-shape hysteresis in \citet{Maitra2004ApJ...608..444M} shows that for a similar count rate, during the rise phase and the decay phase, the source could show different hardness and thus have a different spectral description. But the HID using a softer hardness ratio didn't depict the hysteresis and we observe that the path taken during the rise part of the outburst is similar to the decay part. Although the source showed jumps in hardness at similar counts (e.g. B114, B116, B150 had a higher hardness as compared to B113, B117 and B149) this is different from the traditional hysteresis where the transition from the LHS to HSS and HSS to LHS takes different paths. It was interesting to note that some of these observations have also indicated bursts (table~\ref{tab: Bursts table}). 

\subsection{Spectral and parameter evolution}\label{ssec:par_evol}

The HID of the source indicates a clear evolution of the source throughout its outburst as it traverses through various states. In particular, the source stays for a significant amount of time in the outburst in the HSS, where the source spectrum is adequately described as a combination of a thermalised accretion disc and Compton up-scattering of a fraction of the accretion disc photons. 


\changes{Due to a limited energy range covered by \nicer, the spectral model is found to have strong degeneracies. Mainly, we found that the different combinations of photon index ($\Gamma$) and covering fraction (\textit{f}) of the Comptonising medium result in similar $\chi^2$ statistic (see figure~\ref{fig:contour} for a $\chi^2$ contours for a sample of observations). Although there was a degeneracy between these two parameters, for individual observations the loci of the contours follow different tracks. This indicates, that although these parameters are degenerate, they have a systematic evolution as the source undergoes an outburst. Due to the degeneracy, we investigated the evolution of the parameters by fixing \textit{f} for the duration of the outburst at different values. }

\changes{The choice of the \textit{f} did not affect the estimated value of the absorption column density ($n_{\rm{H}}$) which has a typical value of 0.48--0.51 for almost all observations. This value is higher than the line of sight column density ($=0.31\times10^{22}$~cm$^{-2}$, \citealt{nh2016A&A...594A.116H}) indicating a local absorbing column. 
The $n_{\rm{H}}$ observed is similar to the value reported in the literature \citep{2020MNRAS.492.2858G, li2021ApJ...920...35L}. Some of the observations (B113 -- B117 and B150), however, suggest that the absorption column density is similar to the line of sight value \citep[also seen in \src\ in HXMT observations in a different outburst;][]{2020JHEAp..25...10G}. Notably, these observations also lie on a separate part of HID (see figure~\ref{fig:hid}), occupying a typically harder region for a similar count rate in rest of the observations. This behaviour suggests that the local absorption column is perhaps caused due to a clumpy medium which disappears for a small part of the outburst. And since the local absorbing column variation is independent of assumed covering fraction of the corona, we can safely assume that they are separate entities in the system. 
} 


\changes{The measurement of the temperature of the inner edge of the accretion disc ($T_{\rm{in}}$) depends on the assumption of \textit{f}. A lower covering fraction requires a hotter accretion disc for a similar fit (see figure~\ref{fig: Parameter evolution}). At the same time, for a lower \textit{f}, the normalisation of the accretion disc component ($n_{dbb}$) is also lower. Barring these differences in the value offset, the trends for both parameters of the accretion disc are similar for different \textit{f}. The trends for the accretion disc parameters also conform with each other in different outbursts, indicating a similar behaviour across different epochs. The evolution of $T_{\rm{in}}$ is similar to count rate and hardness evolution i.e. for the observations in HSS, the temperature remains the same. Additionally for the observations in HSS, the disc normalization is observed to be constant indicating a stable disc during the HSS \citep[e.g.][]{Done2007A&ARv..15....1D}. }
The \textit{RXTE} observations have also built a similar picture of the source in the HSS \citep{Maitra2004ApJ...608..444M} but the actual values differ due to the energy range considered and the model assumption used to describe the spectrum. 
As the source moves away from the HSS, the disc temperature reduces and the normalisation (and thus the inner radius) increases.



\changes{As expected from the $\chi^2$ contours, $\Gamma$ has a strong dependence on the assumed covering fraction. But for a given covering fraction, the $\Gamma$ follows the count rate evolution inversely. For a low covering fraction, the $\Gamma$ is consistent with 1.0 (see panels (d) and (i) in figure~\ref{fig: Parameter evolution}, where these points are indicated by triangles), which corresponds to an infinite optical depth of the corona.  
Electron temperature ($kT_{e}$) on the other hand is insensitive to the assumption of a fixed covering fraction (typically varying between 1.5--2~keV) but has a stronger deviation for the observations not in the HSS. }
In the HSS, the parameters of the corona are also roughly constant with minor fluctuations in the electron temperature which is also seen as the jitters in the source position on the HID. \changes{But as the source moves away from the HSS, the spectrum steepens with a simultaneous reduction in the accretion disc temperature and the normalization.  }


 

\changes{Some of the observations indicate residuals near 6.5~keV after modelling the spectrum with Model A. Thus we modelled these residuals with a Gaussian component to account for the Fe K$\alpha$ emission line at 6.5~keV \citep{2003PhR...377..389R}. For example, for observation A116, the emission line can be clearly seen in the residuals plotted in the third panel of \ref{fig:spectral fits}. The observations depicting significant Fe  K$\alpha$ emission are marked with crosses in figures~\ref{fig:lc_hr}, \ref{fig:hid}, and \ref{fig: Parameter evolution}. In all cases of detection, the line width is $\approx0.5$~keV and line normalization is $\approx5\times10^{-3}$ photons cm$^{-2}$ s$^{-1}$. }

\section{Conclusions}

We investigate the spectral properties of \src\ as observed with \nicer\ in \changes{0.5--10~keV} and infer the spectral components present in various canonical states of the source. The extensive \nicer\ observations probed the evolution of the source using the sensitive soft X-ray coverage. The HID constructed using a softer definition of the X-ray colour revealed a track distinct from the hysteresis track typically observed for the source using a hard colour. 
\changes{We find that the spectrum of the source in the HSS can be well described as a partially Comptonised accretion disk. Some of the observations also exhibit reprocessing of the Comptonised emission as an iron line emission. Our analysis is limited by the degeneracy in the parameters of the Comptonising medium, likely due to limited energy range of the instrument. We find evidence for a variation in the local absorption column which also manifests as a separate position on the HID. }

\section*{Acknowledgements}

We express our sincere gratitude to the anonymous referee(s) for their invaluable comments and suggestions, which have significantly improved the quality and clarity of this paper. 
This work was supported by NASA through the \textit{NICER} mission and the Astrophysics Explorers Program. 
This research has also made use of data and/or software provided by the High Energy Astrophysics Science Archive Research Center (HEASARC), which is a service of the Astrophysics Science Division at NASA/GSFC and the High Energy Astrophysics Division of the Smithsonian Astrophysical Observatory. KGP thanks Dr Aru Beri for the help in learning HEASoft.  
\section*{Data Availability}

 The \nicer\ data are available at \url{https://heasarc.gsfc.nasa.gov/docs/archive.html} for public download. The observation IDs mentioned in Tables~\ref{tab: Outburst A log table} and \ref{tab: Outburst B log table} can be used to identify the observations in the archive. 



\bibliographystyle{mnras}
\bibliography{ref} 




\appendix
\section{Tables}

\begin{table}[hbt!]
    \caption{\nicer\ observation log for outburst A. Exposures are net exposure after the standard filtering.}
    \label{tab: Outburst A log table}
    \begin{tabular}{c c c r}
    \hline
    ObsID & Start MJD & Stop MJD & Exposure(ks)\\
    \hline
    2050340102 & 58715.955 & 58715.961 & 0.46\\
    2050340103 & 58716.214 & 58716.242 & 2.44\\
    2050340104 & 58717.246 & 58717.269 & 1.99\\
    2050340105 & 58718.357 & 58718.384 & 2.37\\
    2050340106 & 58719.453 & 58719.459 & 0.52\\
    2050340107 & 58720.297 & 58720.307 & 0.81\\
    2050340108 & 58721.588 & 58721.593 & 0.4\\
    2050340109 & 58722.046 & 58722.069 & 1.99\\
    2050340110 & 58723.327 & 58723.337 & 0.88\\
    2050340111 & 58725.076 & 58725.089 & 1.14\\
    2050340112 & 58726.682 & 58726.686 & 0.41\\
    2050340113 & 58727.074 & 58727.097 & 1.97\\
    2050340114 & 58728.743 & 58728.768 & 2.17\\
    2050340115 & 58729.376 & 58729.397 & 1.82\\
    2050340116 & 58730.169 & 58730.178 & 0.77\\
    2050340117 & 58731.891 & 58731.911 & 1.72\\
    2050340118 & 58732.472 & 58732.483 & 0.92\\
    2050340119 & 58733.117 & 58733.153 & 3.11\\
    2050340120 & 58734.872 & 58734.910 & 3.28\\
    2050340121 & 58734.992 & 58734.029 & 3.2\\
    2050340122 & 58736.986 & 58736.999 & 1.17\\
    2050340123 & 58737.64 & 58737.653 & 1.15\\
    2050340124 & 58738.147 & 58738.165 & 1.54\\
    2050340125 & 58739.439 & 58739.448 & 0.81\\
    2050340126 & 58740.280 & 58740.286 & 0.52\\
    2050340127 & 58743.777 & 58743.789 & 0.99\\
    2050340128 & 58744.423 & 58744.434 & 0.96\\
    2050340129 & 58745.329 & 58745.337 & 0.67\\    
    \hline
        \end{tabular}
\end{table}

\begin{table}[hbt!] 
    \caption{\nicer\ observation log for outburst B. Exposures are net exposure after the standard filtering.}
    \label{tab: Outburst B log table}
    \begin{tabular}{c c c r}
    \hline
    ObsID & Start MJD & Stop MJD & Exposure(ks)\\
    \hline
    3050340108 & 59082.082 & 59082.115 & 2.84\\
    3050340109 & 59083.115 & 59083.153 & 3.29\\
    3050340110 & 59084.147 & 59084.201 & 4.64\\
    3050340111 & 59084.002 & 59084.052 & 4.36\\
    3050340112 & 59086.016 & 59086.071 & 4.75\\
    3050340113 & 59086.002 & 59086.056 & 4.66\\
    3050340114 & 59088.014 & 59088.074 & 5.14\\
    3050340115 & 59089.047 & 59089.111 & 5.54\\
    3050340116 & 59090.016 & 59090.061 & 3.93\\
    3050340117 & 59090.984 & 59091.053 & 5.92\\
    3050340118 & 59092.018 & 59092.073 & 4.79\\
    3050340119 & 59093.003 & 59093.044 & 3.53\\
    3050340120 & 59094.019 & 59094.031 & 1.06\\
    3050340121 & 59095.069 & 59095.117 & 4.17\\
    3050340122 & 59096.217 & 59096.256 & 3.39\\
    3050340123 & 59097.007 & 59097.154 & 12.7\\
    3050340124 & 59098.213 & 59098.248 & 2.98\\
    3050340125 & 59099.991 & 59100.078 & 7.5\\
    3050340126 & 59100.294 & 59100.332 & 3.26\\
    3050340127 & 59101.312 & 59101.345 & 2.87\\
    3050340128 & 59102.216 & 59102.251 & 3.1\\
    3050340129 & 59103.444 & 59103.453 & 0.75\\
    3050340130 & 59104.416 & 59104.424 & 0.75\\
    3050340131 & 59106.363 & 59106.376 & 1.16\\
    3050340132 & 59107.145 & 59107.159 & 1.19\\
    3050340133 & 59108.432 & 59108.461 & 2.55\\
    3050340134 & 59109.146 & 59109.157 & 0.99\\
    3050340135 & 59120.575 & 59120.595 & 1.75\\
    3050340136 & 59121.156 & 59121.183 & 2.32\\
    3050340137 & 59121.992 & 59122.027 & 3.02\\
    3050340138 & 59123.088 & 59123.113 & 2.15\\
    3050340139 & 59124.06 & 59124.082 & 1.91\\
    3050340140 & 59125.026 & 59125.089 & 5.47\\
    3050340141 & 59126.188 & 59126.203 & 1.34\\
    3050340142 & 59127.220 & 59127.245 & 2.17\\
    3050340143 & 59129.608 & 59129.614 & 0.46\\
    3050340144 & 59130.641 & 59130.682 & 3.51\\
    3050340145 & 59131.032 & 59131.27 & 20.58\\
    3050340146 & 59131.995 & 59132.135 & 12.11\\
    3050340147 & 59133.099 & 59133.124 & 2.17\\
    3050340148 & 59134.068 & 59134.071 & 0.31\\
    3050340149 & 59136.379 & 59136.380 & 0.03\\
    3050340150 & 59140.119 & 59140.131 & 1.05\\
    3050340151 & 59142.069 & 59142.083 & 1.21\\
    3050340152 & 59144.266 & 59144.278 & 1.05\\
    3050340153 & 59146.379 & 59146.395 & 1.31\\
    3050340154 & 59148.381 & 59148.394 & 1.17\\
    3050340156 & 59152.449 & 59152.459 & 0.93\\
    3050340157 & 59154.063 & 59154.087 & 2.04\\
    3050340158 & 59155.937 & 59155.948 & 1.01\\
    3050340159 & 59157.164 & 59157.173 & 0.81\\    
    \hline
    \end{tabular}
\end{table}

\begin{table*}[hbt!]
    \caption{Log of all the burst intervals observed in the \nicer\ monitoring of 2019 and 2020 outburst of the source. \changes{The time mentioned here is in spacecraft time. The mean count rate is the mean of the count rate before the outburst. The pean count rate is the peak count rate of the burst. For the observation 3050340116, the decay of a burst is observed, so the peak count rate is uncertain. } %
    }
    \label{tab: Bursts table}
    \begin{tabular}{c c c c c}
    \hline
    ObsID & Time interval (in s) & Duration(s) & Mean count rate & Peak count rate\\
    \hline
    3050340110 & 209658056-209658596 & 540 & 585.85 & 5497.43\\ 
    3050340111 & 209719346-209719646 & 300 & 398.96 & 3237.50\\
    3050340114 & 209998142-209998642 & 500 & 161.68 & 3627.87\\
    3050340116 & 210154101-210154293 & 192 & 130.91 & 186.37 \\
    3050340123 & 210767412-210767912 & 500 & 1857.85 & 2225.94\\ 
    3050340142 & 213346189-213346299 & 110 & 2700.36 & 4324.25\\
    3050340150 & 214455474-214455554 & 80  & 268.22 & 1566.81\\
    3050340150 & 214455934-214455974 & 40  & 246.02 & 1664.31\\
    \hline
        \end{tabular}
        
\end{table*}

\begin{table*}
        \begin{threeparttable}
        \caption{\changes{Outburst A (2019) spectral parameter table for \textit{f}=0.75 assuming the models:  Model A: \texttt{tbabs*(gauss+thcomp$\otimes$diskbb)} and Model B: \texttt{tbabs*(gauss+gauss+thcomp$\otimes$diskbb)}.Errors listed here are $1\sigma$ errors and they are not listed for the observations with $\chi^{2}$/DoF $> 1.5$ and for parameter with error of the order 0.001.}
    }
    \label{tab:2019 Outburst Parameters} 
    \begin{tabular}{||c|c|c c| c c| c c| c|c||}

    \hline
    \multicolumn{1}{|c|}{} &
    \multicolumn{1}{|c|}{} &
    \multicolumn{2}{|c|}{Gaussian (6.5~keV)\tnote{$\alpha$}} &
    \multicolumn{2}{|c|}{diskbb} &
    \multicolumn{2}{|c|}{thcomp} &
    \multicolumn{1}{|c|}{Flux\tnote{$\beta$}}&
    \multicolumn{1}{|c|}{}\\
    \hline
    ID & $n_{H}$\tnote{$\delta$} & Sigma (keV) & norm ($10^{-3}$) &  $T_{\rm in}$ (keV) & $n_{dbb} (10^{2})$ &  $\Gamma$ &  $kT_{\rm e}$ (keV) & $10^{-8}$erg/cm$^2$/s &  $\chi^2$ / DoF \\
    \hline   

A102 & $0.5_{}^{}$ & $-_{}^{}$ & $-_{}^{}$ & $0.71_{-0.01}^{+0.01}$ & $16.69_{-1.1}^{+1.2}$ & $1.64_{-0.01}^{+0.01}$ & $1.85_{-0.02}^{+0.02}$ & $1.869_{-0.004}^{+0.004}$  & 146 / 142\\
A103 & $0.5_{}^{}$ & $-_{}^{}$ & $-_{}^{}$ & $0.72_{-0.01}^{+0.01}$ & $14.98_{-0.6}^{+0.7}$ & $1.7_{-0.01}^{+0.01}$ & $1.98_{-0.02}^{+0.02}$ & $1.74_{-0.002}^{+0.002}$  & 119 / 153\\
A104 & $0.5_{}^{}$ & $0.44_{-0.07}^{+0.08}$ & $5.1_{-1.0}^{+0.8}$ & $0.69_{-0.01}^{+0.01}$ & $17.09_{-0.8}^{+0.8}$ & $1.72_{-0.01}^{+0.01}$ & $1.84_{-0.02}^{+0.02}$ & $1.609_{-0.002}^{+0.002}$  & 147 / 152\\
A105 & $0.49_{}^{}$ & $0.43_{-0.08}^{+0.1}$ & $4.1_{-0.9}^{+0.5}$ & $0.74_{-0.01}^{+0.01}$ & $13.6_{-0.7}^{+0.6}$ & $1.67_{-0.01}^{+0.01}$ & $1.91_{-0.02}^{+0.02}$ & $1.792_{-0.002}^{+0.002}$  & 135 / 154\\
A106 & $0.5_{}^{}$ & $0.5_{}\tnote{*}$ & $6.5_{-1.6}^{+1.6}$ & $0.7_{-0.01}^{+0.01}$ & $17.2_{-1.1}^{+1.2}$ & $1.64_{-0.01}^{+0.01}$ & $1.85_{-0.02}^{+0.03}$ & $1.822_{-0.004}^{+0.004}$  & 131 / 145\\
A107 & $0.5_{}^{}$ & $0.52_{-0.14}^{+0.25}$ & $4.9_{-1.6}^{+2.5}$ & $0.7_{-0.01}^{+0.01}$ & $16.04_{-0.9}^{+1.0}$ & $1.67_{-0.01}^{+0.01}$ & $1.87_{-0.02}^{+0.02}$ & $1.69_{-0.003}^{+0.003}$  & 154 / 147\\
A108 & $0.49_{}^{}$ & $0.65_{-0.15}^{+0.19}$ & $7.8_{-2.3}^{+3.1}$ & $0.68_{-0.01}^{+0.02}$ & $15.72_{-1.1}^{+1.2}$ & $1.78_{-0.01}^{+0.02}$ & $1.78_{-0.03}^{+0.04}$ & $1.366_{-0.003}^{+0.003}$  & 137 / 137\\
A109 & $0.5_{}^{}$ & $0.62_{-0.1}^{+0.11}$ & $7.1_{-1.4}^{+1.6}$ & $0.7_{-0.01}^{+0.01}$ & $16.11_{-0.8}^{+0.8}$ & $1.72_{-0.01}^{+0.01}$ & $1.82_{-0.02}^{+0.02}$ & $1.611_{-0.002}^{+0.002}$  & 139 / 152\\
A110 & $0.5_{}^{}$ & $0.7_{-0.17}^{+0.21}$ & $6.0_{-2.0}^{+2.7}$ & $0.74_{-0.01}^{+0.01}$ & $13.28_{-0.8}^{+0.8}$ & $1.69_{-0.01}^{+0.01}$ & $1.9_{-0.02}^{+0.03}$ & $1.728_{-0.003}^{+0.003}$  & 112 / 148\\
A111 & $0.5_{}^{}$ & $0.45_{-0.1}^{+0.11}$ & $4.6_{-1.2}^{+1.3}$ & $0.69_{-0.01}^{+0.01}$ & $16.93_{-0.9}^{+1.0}$ & $1.7_{-0.01}^{+0.01}$ & $1.8_{-0.02}^{+0.02}$ & $1.614_{-0.003}^{+0.003}$  & 162 / 150\\
A112 & $0.49_{}^{}$ & $-_{}^{}$ & $-_{}^{}$ & $0.71_{-0.02}^{+0.02}$ & $15.51_{-1.1}^{+1.2}$ & $1.69_{-0.01}^{+0.01}$ & $1.85_{-0.03}^{+0.03}$ & $1.69_{-0.004}^{+0.004}$  & 150 / 141\\
A113 & $0.49_{}^{}$ & $0.42_{-0.06}^{+0.07}$ & $5.2_{-0.9}^{+1.0}$ & $0.72_{-0.01}^{+0.01}$ & $14.93_{-0.7}^{+0.7}$ & $1.71_{-0.01}^{+0.01}$ & $1.91_{-0.02}^{+0.02}$ & $1.679_{-0.002}^{+0.002}$  & 138 / 152\\
A114 & $0.49_{}^{}$ & $0.64_{-0.13}^{+0.15}$ & $6.8_{-1.6}^{+2.0}$ & $0.75_{-0.01}^{+0.01}$ & $13.25_{-0.6}^{+0.6}$ & $1.7_{-0.01}^{+0.01}$ & $1.98_{-0.02}^{+0.02}$ & $1.754_{-0.002}^{+0.002}$  & 121 / 153\\
A115 & $0.5_{}^{}$ & $-_{}^{}$ & $-_{}^{}$ & $0.72_{-0.01}^{+0.01}$ & $14.86_{-0.6}^{+0.7}$ & $1.69_{-0.01}^{+0.01}$ & $1.92_{-0.02}^{+0.02}$ & $1.735_{-0.002}^{+0.002}$  & 184 / 154\\
A116 & $0.5_{}^{}$ & $0.61_{-0.12}^{+0.14}$ & $8.9_{-2.0}^{+2.5}$ & $0.68_{-0.01}^{+0.01}$ & $16.7_{-1.0}^{+1.0}$ & $1.74_{-0.01}^{+0.01}$ & $1.84_{-0.02}^{+0.03}$ & $1.52_{-0.003}^{+0.003}$  & 133 / 145\\
A117 & $0.49_{}^{}$ & $0.54_{-0.11}^{+0.13}$ & $4.7_{-1.2}^{+1.4}$ & $0.74_{-0.01}^{+0.01}$ & $12.77_{-0.6}^{+0.6}$ & $1.7_{-0.01}^{+0.01}$ & $1.97_{-0.02}^{+0.02}$ & $1.642_{-0.002}^{+0.002}$  & 97 / 152\\
A118 & $0.49_{}^{}$ & $0.42_{-0.14}^{+0.18}$ & $3.9_{-1.3}^{+1.8}$ & $0.73_{-0.01}^{+0.01}$ & $13.49_{-0.8}^{+0.8}$ & $1.71_{-0.01}^{+0.01}$ & $1.95_{-0.03}^{+0.03}$ & $1.633_{-0.003}^{+0.003}$  & 183 / 150\\
A119 & $0.5_{}^{}$ & $-_{}^{}$ & $-_{}^{}$ & $0.69_{-0.01}^{+0.01}$ & $15.79_{-0.7}^{+0.5}$ & $1.7_{-0.01}^{+0.01}$ & $1.89_{-0.01}^{+0.01}$ & $1.522_{-0.003}^{+0.003}$  & 180 / 154\\
A120 & $0.5_{}^{}$ & $0.51_{-0.07}^{+0.08}$ & $5.5_{-0.9}^{+1.0}$ & $0.7_{-0.01}^{+0.01}$ & $14.48_{-0.6}^{+0.7}$ & $1.73_{-0.01}^{+0.01}$ & $1.87_{-0.02}^{+0.02}$ & $1.436_{-0.003}^{+0.003}$  & 107 / 153\\
A121 & $0.49_{}^{}$ & $0.62_{-0.13}^{+0.15}$ & $4.3_{-1.1}^{+1.3}$ & $0.72_{-0.01}^{+0.01}$ & $12.95_{-0.6}^{+0.6}$ & $1.73_{-0.01}^{+0.01}$ & $1.92_{-0.02}^{+0.02}$ & $1.445_{-0.002}^{+0.002}$  & 114 / 154\\
A122 & $0.49_{}^{}$ & $0.02_{-0.02}^{+0.08}$ & $0.6_{-0.3}^{+0.3}$ & $0.75_{-0.02}^{+0.01}$ & $10.52_{-0.5}^{+0.5}$ & $1.77_{-0.01}^{+0.01}$ & $2.13_{-0.03}^{+0.03}$ & $1.365_{-0.002}^{+0.002}$  & 114 / 149\\
A123 & $0.5_{}^{}$ & $0.69_{-0.18}^{+0.25}$ & $4.4_{-1.6}^{+2.2}$ & $0.71_{-0.01}^{+0.01}$ & $12.25_{-0.6}^{+0.6}$ & $1.77_{-0.01}^{+0.01}$ & $1.93_{-0.03}^{+0.03}$ & $1.237_{-0.002}^{+0.002}$  & 132 / 148\\
A124 & $0.49_{}^{}$ & $0.7_{-0.11}^{+0.13}$ & $6.0_{-1.3}^{+1.6}$ & $0.72_{-0.01}^{+0.01}$ & $10.7_{-0.5}^{+0.6}$ & $1.85_{-0.01}^{+0.01}$ & $1.97_{-0.03}^{+0.03}$ & $1.109_{-0.002}^{+0.002}$  & 116 / 148\\
A125 & $0.5_{}^{}$ & $-_{}^{}$ & $-_{}^{}$ & $0.7_{-0.01}^{+0.01}$ & $10.97_{-0.5}^{+0.6}$ & $1.91_{-0.01}^{+0.01}$ & $2.2_{-0.05}^{+0.06}$ & $1.001_{-0.002}^{+0.002}$  & 158 / 141\\
A126 & $0.5_{}^{}$ & $-_{}^{}$ & $-_{}^{}$ & $0.69_{-0.01}^{+0.01}$ & $10.89_{-0.7}^{+0.7}$ & $1.97_{-0.02}^{+0.02}$ & $2.19_{-0.07}^{+0.08}$ & $0.875_{-0.003}^{+0.003}$  & 130 / 136\\
A127 & $0.51_{}^{}$ & $-_{}^{}$ & $-_{}^{}$ & $0.58_{-0.01}^{+0.01}$ & $12.95_{-0.6}^{+0.7}$ & $2.2_{-0.02}^{+0.02}$ & $3.28_{-0.22}^{+0.3}$ & $0.469_{-0.001}^{+0.001}$  & 183 / 137\\
A128\tnote{$\dagger$} & $0.5_{}^{}$ & $-_{}^{}$ & $-_{}^{}$ & $0.49_{}^{}$ & $16.7_{}^{}$ & $1.98_{}^{}$ & $131.1_{}^{}$ & $0.303_{}^{}$  & 487 / 135\\
A129\tnote{$\dagger$} & $0.52_{}^{}$ & $-_{}^{}$ & $-_{}^{}$ & $0.35_{}^{}$ & $42.37_{}^{}$ & $2.24_{}^{}$ & $3.5_{}^{}$ & $0.22_{}^{}$  & 373 / 126\\
\hline

    \end{tabular}
    \begin{tablenotes}
         \item [*] frozen to the values.
         \item[${\dagger}$] observations with $\chi^{2}$/DoF$>$1.5.
        \item [$\delta$] The 1$\sigma$ errors for $n_{H}$ are $\approx$ 0.002.
        \item [$\alpha$] The Gaussian line energy for Fe $k\alpha$ line is fixed at 6.5~keV.
        \item [$\beta$] The flux is the unabsorbed flux computed in the energy range 0.5--12~keV.
        \item [$\gamma$] The instrumental Gaussian is present in all the observations at around 0.75~keV with a width of   $\approx$ 0.04~keV.
        
    \end{tablenotes}
   \end{threeparttable}
\end{table*}

\begin{table*}
    \scalebox{0.91}{
    \begin{minipage}{\textwidth}
    
    \begin{threeparttable}
    \caption{\changes{Outburst B (2020) spectral parameter table for \textit{f}=0.75.  assuming the models: Model A: \texttt{tbabs*(gauss+thcomp$\otimes$diskbb)}, Model B: \texttt{tbabs*(gauss+gauss+thcomp$\otimes$diskbb)}, Model C: \texttt{tbabs*(thcomp$\otimes$diskbb)}, Model D: \texttt{tbabs*(diskbb)}. Rest of the configuration is same as table~\ref{tab:2019 Outburst Parameters}. 
    }}
    \label{tab:2020 Outburst Parameters} 
    \begin{tabular}{||c|c|c c| c c| c c| c|c||}

    \hline
    \multicolumn{1}{|c|}{} &
    \multicolumn{1}{|c|}{} &
    \multicolumn{2}{|c|}{Gaussian (6.5~keV)\tnote{$\alpha$}} &
    \multicolumn{2}{|c|}{diskbb} &
    \multicolumn{2}{|c|}{thcomp} &
    \multicolumn{1}{|c|}{Flux\tnote{$\beta$}}&
    \multicolumn{1}{|c|}{}\\
    \hline
    ID & $n_{H}$\tnote{$\delta$} & Sigma (keV) & norm ($10^{-3}$) &  $T_{\rm in}$ (keV) & $n_{dbb} (10^{2})$ &  $\Gamma$ &  $kT_{\rm e}$ (keV) &(10$^{-8}$ erg/cm$^2$/s) &  $\chi^2$ / DoF \\
    \hline 
    
B108\tnote{$\dagger$} & $0.48_{}^{}$ & $-_{}^{}$ & $-_{}^{}$ & $0.38_{}^{}$ & $20.39_{}^{}$ & $2.03_{}^{}$ & $79.25_{}^{}$ & $0.135_{}^{}$  & 786 / 137\\
B109\tnote{$\dagger$} & $0.5_{}^{}$ & $-_{}^{}$ & $-_{}^{}$ & $0.46_{}^{}$ & $15.81_{}^{}$ & $2.04_{}^{}$ & $150.0_{}^{}$ & $0.217_{}^{}$  & 268 / 140\\
B110\tnote{$\dagger$} & $0.5_{}^{}$ & $-_{}^{}$ & $-_{}^{}$ & $0.49_{}^{}$ & $13.34_{}^{}$ & $2.1_{}^{}$ & $119.97_{}^{}$ & $0.229_{}^{}$  & 467 / 142\\
B111\tnote{$\dagger$} & $0.49_{}^{}$ & $1.1_{}^{}$ & $2.4_{}^{}$ & $0.42_{}^{}$ & $17.94_{}^{}$ & $2.14_{}^{}$ & $71.42_{}^{}$ & $0.161_{}^{}$  & 769 / 140\\
B112\tnote{$\dagger$} & $0.5_{}^{}$ & $-_{}^{}$ & $-_{}^{}$ & $0.48_{}^{}$ & $12.63_{}^{}$ & $2.14_{}^{}$ & $108.74_{}^{}$ & $0.189_{}^{}$  & 453 / 140\\
B113\tnote{$\dagger$} & $0.46_{}^{}$ & $-_{}^{}$ & $-_{}^{}$ & $0.42_{}^{}$ & $11.25_{}^{}$ & $1.95_{}^{}$ & $114.46_{}^{}$ & $0.119_{}^{}$  & 219 / 138\\
B114\tnote{$\dagger$} & $0.36_{}^{}$ & $-_{}^{}$ & $-_{}^{}$ & $0.53_{}^{}$ & $2.04_{}^{}$ & $1.81_{}^{}$ & $3.55_{}^{}$ & $0.072_{}^{}$  & 233 / 137\\
B115 & $0.34_{}^{}$ & $-_{}^{}$ & $-_{}^{}$ & $0.57_{-0.01}^{+0.01}$ & $1.4_{-0.1}^{+0.1}$ & $1.84_{-0.01}^{+0.01}$ & $4.36_{-0.45}^{+0.78}$ & $0.064_{}^{}$  & 175 / 135\\
B116 & $0.37_{}^{}$ & $-_{}^{}$ & $-_{}^{}$ & $0.67_{-0.01}^{+0.01}$ & $0.74_{}^{}$ & $1.82_{-0.01}^{+0.01}$ & $3.5_{}\tnote{*}$ & $0.066_{}^{}$  & 163 / 134\\
B117 & $0.43_{-0.01}^{+0.01}$ & $0.5_{}\tnote{*}$ & $0.3_{-0.1}^{+0.1}$ & $0.34_{-0.02}^{+0.01}$ & $21.47_{-2.9}^{+4.7}$ & $1.74_{-0.01}^{+0.01}$ & $3.16_{-0.08}^{+0.09}$ & $0.155_{}^{}$  & 153 / 144\\
B118\tnote{$\dagger$} & $0.49_{}^{}$ & $0.5_{}\tnote{*}$ & $2.0_{}^{}$ & $0.44_{}^{}$ & $25.36_{}^{}$ & $1.94_{}^{}$ & $53.37_{}^{}$ & $0.332_{}^{}$  & 281 / 148\\
B119 & $0.5_{}^{}$ & $0.5_{}\tnote{*}$ & $2.5_{-0.3}^{+0.3}$ & $0.59_{-0.01}^{+0.01}$ & $14.32_{-0.5}^{+0.4}$ & $2.17_{-0.01}^{+0.01}$ & $2.94_{-0.1}^{+0.12}$ & $0.567_{-0.001}^{+0.001}$  & 181 / 150\\
B120 & $0.5_{}^{}$ & $0.5_{}\tnote{*}$ & $3.7_{-0.5}^{+0.5}$ & $0.67_{-0.01}^{+0.01}$ & $10.1_{-0.4}^{+0.5}$ & $2.22_{-0.02}^{+0.02}$ & $2.86_{-0.17}^{+0.21}$ & $0.665_{-0.002}^{+0.002}$  & 106 / 139\\
B121 & $0.5_{}^{}$ & $-_{}^{}$ & $-_{}^{}$ & $0.69_{-0.01}^{+0.01}$ & $9.47_{-0.3}^{+0.4}$ & $1.91_{-0.01}^{+0.01}$ & $2.14_{-0.03}^{+0.03}$ & $0.813_{-0.001}^{+0.001}$  & 145 / 153\\
B122 & $0.5_{}^{}$ & $-_{}^{}$ & $-_{}^{}$ & $0.67_{-0.01}^{+0.01}$ & $10.75_{-0.4}^{+0.4}$ & $1.88_{-0.01}^{+0.01}$ & $2.02_{-0.03}^{+0.02}$ & $0.815_{-0.001}^{+0.001}$  & 149 / 153\\
B123 & $0.5_{}^{}$ & $0.56_{-0.07}^{+0.08}$ & $3.3_{-0.5}^{+0.5}$ & $0.68_{-0.01}^{+0.01}$ & $10.91_{-0.4}^{+0.4}$ & $1.83_{-0.01}^{+0.01}$ & $1.93_{-0.01}^{+0.01}$ & $0.933_{}^{}$  & 117 / 167\\
B124 & $0.5_{}^{}$ & $0.33_{-0.08}^{+0.1}$ & $2.2_{-0.5}^{+0.6}$ & $0.7_{-0.01}^{+0.01}$ & $11.4_{-0.5}^{+0.5}$ & $1.82_{-0.01}^{+0.01}$ & $1.94_{-0.02}^{+0.02}$ & $1.091_{-0.001}^{+0.001}$  & 93 / 152\\
B125 & $0.5_{}^{}$ & $0.47_{-0.1}^{+0.07}$ & $3.8_{-0.6}^{+0.6}$ & $0.69_{-0.01}^{+0.01}$ & $13.32_{-0.5}^{+0.5}$ & $1.79_{-0.01}^{+0.01}$ & $1.92_{-0.01}^{+0.01}$ & $1.229_{}^{}$  & 109 / 161\\
B126 & $0.5_{}^{}$ & $0.54_{-0.11}^{+0.14}$ & $2.6_{-0.7}^{+0.9}$ & $0.69_{-0.01}^{+0.01}$ & $15.06_{-0.6}^{+0.7}$ & $1.75_{-0.01}^{+0.01}$ & $1.86_{-0.02}^{+0.01}$ & $1.373_{}^{}$  & 101 / 152\\
B127 & $0.49_{}^{}$ & $0.35_{-0.07}^{+0.08}$ & $3.1_{-0.7}^{+0.7}$ & $0.72_{-0.01}^{+0.01}$ & $13.65_{-0.6}^{+0.6}$ & $1.7_{-0.01}^{+0.01}$ & $1.9_{-0.02}^{+0.02}$ & $1.562_{-0.002}^{+0.002}$  & 114 / 153\\
B128 & $0.5_{}^{}$ & $0.62_{-0.15}^{+0.2}$ & $4.7_{-1.3}^{+1.9}$ & $0.71_{-0.01}^{+0.01}$ & $14.7_{-0.7}^{+0.7}$ & $1.69_{-0.01}^{+0.01}$ & $1.88_{-0.01}^{+0.01}$ & $1.604_{}^{}$  & 114 / 153\\
B129 & $0.49_{}^{}$ & $0.56_{-0.18}^{+0.28}$ & $4.2_{-1.7}^{+2.5}$ & $0.72_{-0.01}^{+0.01}$ & $13.91_{-0.9}^{+0.9}$ & $1.69_{-0.01}^{+0.01}$ & $1.87_{-0.02}^{+0.03}$ & $1.603_{-0.002}^{+0.003}$  & 135 / 145\\
B130 & $0.5_{}^{}$ & $-_{}^{}$ & $-_{}^{}$ & $0.66_{-0.01}^{+0.01}$ & $18.21_{-1.1}^{+1.1}$ & $1.71_{-0.01}^{+0.01}$ & $1.85_{-0.01}^{+0.02}$ & $1.464_{-0.003}^{+0.003}$  & 148 / 145\\
B131 & $0.5_{}^{}$ & $-_{}^{}$ & $-_{}^{}$ & $0.71_{-0.01}^{+0.01}$ & $13.71_{-0.7}^{+0.7}$ & $1.71_{-0.01}^{+0.01}$ & $1.99_{-0.02}^{+0.02}$ & $1.529_{-0.003}^{+0.003}$  & 126 / 150\\
B132 & $0.5_{}^{}$ & $-_{}^{}$ & $-_{}^{}$ & $0.71_{-0.01}^{+0.01}$ & $14.45_{-0.7}^{+0.8}$ & $1.68_{-0.01}^{+0.01}$ & $1.95_{-0.02}^{+0.02}$ & $1.582_{-0.003}^{+0.003}$  & 132 / 151\\
B133 & $0.5_{}^{}$ & $-_{}^{}$ & $-_{}^{}$ & $0.7_{-0.01}^{+0.01}$ & $14.23_{-0.6}^{+0.6}$ & $1.74_{-0.01}^{+0.01}$ & $1.92_{-0.02}^{+0.02}$ & $1.402_{-0.002}^{+0.002}$  & 129 / 153\\
B134 & $0.49_{}^{}$ & $0.63_{-0.11}^{+0.13}$ & $7.5_{-1.7}^{+2.0}$ & $0.72_{-0.01}^{+0.01}$ & $12.56_{-0.7}^{+0.7}$ & $1.74_{-0.01}^{+0.01}$ & $1.9_{-0.03}^{+0.03}$ & $1.395_{-0.003}^{+0.003}$  & 109 / 147\\
B135 & $0.49_{}^{}$ & $0.61_{-0.25}^{+0.22}$ & $4.1_{-1.6}^{+1.9}$ & $0.72_{-0.01}^{+0.01}$ & $12.74_{-0.6}^{+0.7}$ & $1.71_{-0.01}^{+0.01}$ & $1.98_{-0.02}^{+0.02}$ & $1.446_{-0.002}^{+0.002}$  & 118 / 151\\
B136 & $0.49_{}^{}$ & $0.21_{-0.05}^{+0.07}$ & $2.2_{-0.5}^{+0.6}$ & $0.7_{-0.01}^{+0.01}$ & $13.7_{-0.6}^{+0.6}$ & $1.73_{-0.01}^{+0.01}$ & $1.89_{-0.02}^{+0.02}$ & $1.374_{-0.002}^{+0.001}$  & 114 / 152\\
B137 & $0.5_{}^{}$ & $0.56_{-0.09}^{+0.1}$ & $4.5_{-0.9}^{+1.0}$ & $0.69_{-0.01}^{+0.01}$ & $14.22_{-0.6}^{+0.7}$ & $1.73_{-0.01}^{+0.01}$ & $1.85_{-0.02}^{+0.02}$ & $1.336_{-0.002}^{+0.002}$  & 122 / 153\\
B138 & $0.49_{}^{}$ & $0.49_{-0.1}^{+0.12}$ & $3.6_{-0.9}^{+1.0}$ & $0.71_{-0.01}^{+0.01}$ & $12.59_{-0.6}^{+0.6}$ & $1.78_{-0.01}^{+0.01}$ & $1.97_{-0.02}^{+0.02}$ & $1.299_{-0.002}^{+0.002}$  & 110 / 152\\
B139 & $0.5_{}^{}$ & $0.23_{-0.05}^{+0.08}$ & $1.8_{-0.5}^{+0.6}$ & $0.71_{-0.01}^{+0.01}$ & $12.59_{-0.6}^{+0.6}$ & $1.72_{-0.01}^{+0.01}$ & $1.95_{-0.02}^{+0.02}$ & $1.384_{-0.002}^{+0.002}$  & 141 / 152\\
B140 & $0.5_{}^{}$ & $0.58_{-0.05}^{+0.11}$ & $4.2_{-0.9}^{+1.0}$ & $0.71_{-0.01}^{+0.01}$ & $12.52_{-0.5}^{+0.4}$ & $1.74_{-0.01}^{+0.01}$ & $1.91_{-0.01}^{+0.02}$ & $1.332_{}^{}$  & 102 / 156\\
B141 & $0.49_{}^{}$ & $0.48_{-0.1}^{+0.15}$ & $4.6_{-1.1}^{+1.5}$ & $0.73_{-0.01}^{+0.01}$ & $11.41_{-0.6}^{+0.6}$ & $1.74_{-0.01}^{+0.01}$ & $2.0_{-0.03}^{+0.03}$ & $1.344_{-0.002}^{+0.002}$  & 124 / 149\\
B142 & $0.49_{}^{}$ & $0.65_{-0.21}^{+0.27}$ & $4.5_{-1.6}^{+2.3}$ & $0.73_{-0.01}^{+0.01}$ & $11.23_{-0.6}^{+0.6}$ & $1.76_{-0.01}^{+0.01}$ & $2.0_{-0.02}^{+0.02}$ & $1.286_{-0.002}^{+0.002}$  & 138 / 152\\
B143 & $0.5_{}^{}$ & $-_{}^{}$ & $-_{}^{}$ & $0.67_{-0.01}^{+0.01}$ & $14.32_{-1.0}^{+1.1}$ & $1.74_{-0.01}^{+0.01}$ & $1.9_{-0.03}^{+0.03}$ & $1.183_{-0.003}^{+0.003}$  & 132 / 138\\
B144 & $0.49_{}^{}$ & $-_{}^{}$ & $-_{}^{}$ & $0.7_{-0.01}^{+0.01}$ & $12.03_{-0.8}^{+0.7}$ & $1.55_{-0.01}^{+0.01}$ & $1.51_{-0.01}^{+0.01}$ & $1.363_{-0.001}^{+0.001}$  & 191 / 155\\
B145 & $0.49_{}^{}$ & $0.56_{-0.06}^{+0.07}$ & $3.3_{-0.5}^{+0.5}$ & $0.76_{-0.01}^{+0.01}$ & $8.81_{-0.3}^{+0.3}$ & $1.83_{-0.01}^{+0.01}$ & $1.96_{-0.01}^{+0.01}$ & $1.089_{}^{}$  & 125 / 172\\
B146 & $0.49_{}^{}$ & $0.69_{-0.07}^{+0.07}$ & $4.7_{-0.6}^{+0.7}$ & $0.71_{-0.01}^{+0.01}$ & $9.76_{-0.3}^{+0.2}$ & $1.87_{-0.01}^{+0.01}$ & $2.01_{-0.02}^{+0.02}$ & $0.959_{}^{}$  & 122 / 167\\
B147 & $0.49_{}^{}$ & $0.75_{-0.16}^{+0.17}$ & $4.7_{-1.2}^{+1.5}$ & $0.68_{-0.01}^{+0.01}$ & $10.74_{-0.5}^{+0.2}$ & $1.94_{-0.01}^{+0.01}$ & $2.06_{-0.04}^{+0.04}$ & $0.814_{-0.001}^{+0.001}$  & 108 / 149\\
B148 & $0.5_{}^{}$ & $-_{}^{}$ & $-_{}^{}$ & $0.69_{-0.01}^{+0.01}$ & $9.78_{-0.7}^{+0.8}$ & $2.14_{-0.03}^{+0.03}$ & $2.52_{-0.18}^{+0.22}$ & $0.708_{}^{}$  & 149 / 130\\
B150 & $0.39_{-0.01}^{+0.01}$ & $-_{}^{}$ & $-_{}^{}$ & $0.49_{-0.02}^{+0.02}$ & $4.66_{-0.5}^{+0.6}$ & $1.94_{-0.02}^{+0.02}$ & $2.51_{-0.16}^{+0.21}$ & $0.105_{}^{}$  & 121 / 119\\
B151\tnote{$\dagger$} & $0.58_{}^{}$ & $-_{}^{}$ & $-_{}^{}$ & $0.58_{}^{}$ & $0.77_{}^{}$ & $2.06_{}^{}$ & $2.5_{}^{}$ & $0.031_{}^{}$  & 180 / 117\\
B152\tnote{$\dagger$} & $0.66_{}^{}$ & $-_{}^{}$ & $-_{}^{}$ & $0.38_{}^{}$ & $1.26_{}^{}$ & $2.09_{}^{}$ & $2.0_{}^{}$ & $0.013_{}^{}$  & 319 / 107\\
B153 & $0.75_{-0.04}^{+0.04}$ & $-_{}^{}$ & $-_{}^{}$ & $0.33_{-0.04}^{+0.03}$ & $1.61_{-0.5}^{+1.3}$ & $2.27_{-0.03}^{+0.04}$ & $2.0_{}^{}$ & $0.006_{}^{}$  & 123 / 96\\
B154 & $0.5_{-0.03}^{+0.03}$ & $-_{}^{}$ & $-_{}^{}$ & $0.49_{-0.02}^{+0.02}$ & $0.16_{-0.003}^{+0.003}$ & $-_{}^{}$ & $-_{}^{}$ & $0.001_{}^{}$  & 38 / 41\\
B156 & $0.64_{-0.05}^{+0.06}$ & $-_{}^{}$ & $-_{}^{}$ & $0.4_{-0.03}^{+0.03}$ & $0.23_{-0.1}^{+0.1}$ & $-_{}^{}$ & $-_{}^{}$ & $0.001_{}^{}$  & 22 / 25\\
B157 & $0.56_{-0.04}^{+0.05}$ & $-_{}^{}$ & $-_{}^{}$ & $0.41_{-0.02}^{+0.02}$ & $0.17_{}^{+0.1}$ & $-_{}^{}$ & $-_{}^{}$ & $0.001_{}^{}$  & 17 / 22\\
B158 & $0.52_{-0.07}^{+0.08}$ & $-_{}^{}$ & $-_{}^{}$ & $0.41_{-0.03}^{+0.04}$ & $0.13_{}^{+0.1}$ & $-_{}^{}$ & $-_{}^{}$ & $0.001_{}^{}$  & 31 / 24\\
B159 & $0.71_{-0.11}^{+0.13}$ & $-_{}^{}$ & $-_{}^{}$ & $0.32_{-0.04}^{+0.04}$ & $0.57_{-0.3}^{+0.8}$ & $-_{}^{}$ & $-_{}^{}$ & $0.001_{}^{}$  & 14 / 16\\
\hline
    \end{tabular}
    \begin{tablenotes}
        \item [*] frozen to the values. 
        \item[${\dagger}$] observations with $\chi^{2}$/DoF$>$1.5.
        \item [$\delta$] The 1$\sigma$ errors for $n_{H}$ are $\approx$ 0.002.
        \item [$\alpha$] The Gaussian line energy for Fe $k\alpha$ line is fixed at 6.5~keV.
        \item [$\beta$] The flux is the unabsorbed flux computed in the energy range 0.5--12~keV.
        \item [$\gamma$] The instrumental Gaussian is present in all the observations (except B151-B159) at around 0.75~keV with a width of   $\approx$ 0.04~keV.
                
    \end{tablenotes}
    \end{threeparttable}
    \end{minipage}}
\end{table*}


\bsp	
\label{lastpage}
\end{document}